\documentclass[sigconf]{acmart}

\renewcommand\footnotetextcopyrightpermission[1]{} 

\AtBeginDocument{%
  \providecommand\BibTeX{{%
    \normalfont B\kern-0.5em{\scshape i\kern-0.25em b}\kern-0.8em\TeX}}}

\copyrightyear{2021} 
\acmYear{2021} 
\setcopyright{acmlicensed}
\acmConference[CIKM '21]{CIKM}{2021}{Australia}

\usepackage{subfig}
\usepackage{multirow}
\usepackage[inline]{enumitem}

\newcommand{\partitle}[1]{\vspace{1mm}\noindent\textbf{#1}}

\begin{document}

\title[BERT for Target Apps Selection: Analyzing the Diversity and Performance of BERT in Unified Mobile Search]{BERT for Target Apps Selection: Analyzing the Diversity and Performance of BERT in Unified Mobile Search}

\author{Negin Ghasemi}
\email{n.ghasemi@cs.ru.nl}
\affiliation{%
  \institution{Radboud University}
  \city{Nijmegen}
  \country{The Netherlands}
}

\author{Mohammad Aliannejadi}
\email{m.aliannejadi@uva.nl}
\affiliation{%
  \institution{University of Amsterdam}
  \city{Amsterdam}
  \country{The Netherlands}
}

\author{Djoerd Hiemstra}
\email{hiemstra@cs.ru.nl}
\affiliation{%
  \institution{Radboud University}
  \city{Nijmegen}
  \country{The Netherlands}
}

\renewcommand{\shortauthors}{Ghassemi, et al.}

\fancyhead{}

\begin{abstract}
A unified mobile search framework aims to identify the mobile apps that can satisfy a user's information need and route the user's query to them. Previous work has shown that resource descriptions for mobile apps are sparse as they rely on the app's previous queries. This problem puts certain apps in dominance and leaves out the resource-scarce apps from the top ranks. In this case, we need a ranker that goes beyond simple lexical matching. Therefore, our goal is to study the extent of a BERT-based ranker's ability to improve the quality and diversity of app selection. To this end, we compare the results of the BERT-based ranker with other information retrieval models, focusing on the analysis of selected apps diversification. Our analysis shows that the BERT-based ranker selects more diverse apps while improving the quality of baseline results by selecting the relevant apps such as Facebook and Contacts for more personal queries and decreasing the bias towards the dominant resources such as the Google Search app.
\end{abstract}

\keywords{unified mobile search, resource selection, target apps selection}

\maketitle

\section{Introduction}
\label{sec:introduction}
The increasing popularity of intelligent assistants (e.g., Apple Siri, Microsoft Cortana, and Google Assistant) on smartphones has led to novel information-seeking behavior and interactions between users and their devices. The searching format changed from submitting queries to a search engine via a search browser app to using a universal voice-based search interface. Previous work has studied how users access information through a common channel and highlighted the need for a unified mobile search framework that identifies the most useful target apps based on the user's query among a variety of apps~\citep{UniMobile}. The framework would then route the query to the target apps, and the results would be presented as an integrated list.

The study of unified mobile search is very similar to mobile information retrieval (IR), enabling users to perform every IR task on a single mobile device ~\citep{DBLP:series/sbcs/CrestaniMS17}. The first step towards designing a unified mobile search framework is identifying the target apps for a given query, called the target apps selection task ~\citep{UniMobile}. The target apps selection task is similar to the resource selection task in federated search ~\cite{Shokouhi}. However, most federated search systems are built with uncooperative resources with homogeneous data. ~\citet{CallanConnell} proposed a query-based sampling approach to probe uncooperative resources. However, given the heterogeneous nature of a mobile device environment, conventional resource selection approaches such as query probing are impractical.

Moreover, unlike aggregated search systems~\cite{DBLP:conf/sigir/ArguelloDP10,DBLP:conf/cikm/ArguelloDC11} where the data is heterogeneous, but resources are cooperative, the heterogeneity of data, together with uncooperativeness of resources (i.e., mobile apps), makes the task of target apps selection challenging. Prior work shows that resource descriptions for a mobile app can be created based on the app's previous queries ~\cite{UniMobile, Situ}. However, there is a big difference between the existing amount of data for each app. On the one hand, there are general and popular apps with large number of previous queries. On the other hand, there are more specific apps with a few number of queries. Therefore, creating resource descriptions for apps may lead to sparse representations for unpopular apps; hence the target apps selection may lead to biased results in favor of more popular apps.

Prior work ~\cite{UniMobile,DBLP:journals/corr/abs-2101-03394} has emphasized the significance of modeling target apps selection as query classification. ~\citet{Park} suggested that users' reviews can be used to find out apps' most valuable features and model app descriptions. In a more recent work, another model was proposed by ~\citet{UniMobile} to address the sparsity issue by using a neural model approach that learns a description for each app. Finally, this work was extended by ~\citet{Situ}, using users' contextual information.

To avoid the sparsity limitation and its consequences, it is critical for the target apps selection system to model both query structure and semantics. Recently, transformers-based pre-trained language representations have been counted as a promising approach to model the language's lexical structure and semantics. For example, BERT has been successfully used in ranking tasks, achieving promising results ~\citep{passageReRanking, MultiStage, CEDR}. 
We model the target apps selection task as a ranking problem where we represent uncooperative resources using a BERT-based ranker.

In our work, we aim to analyze BERT ~\citep{BERT} for target apps selection, understanding the extent that this model can take user query characteristics and structure into account. We hypothesize that BERT's knowledge of the natural language, together with its ability to model the semantics of queries can be useful for target apps selection with limited knowledge of the app representations. In such cases, understanding the nature and structure of queries is critical. For instance, queries submitted to Google Search are closer to natural language, whereas queries submitted to Contacts are keyword-based and often person names.

To this aim, we fine-tune BERT for the task of target app selection on the UniMobile dataset introduced by ~\citet{UniMobile}, addressing the following research questions:
\vspace{1mm}
\begin{itemize}[leftmargin=*]
\item \textbf{RQ1}: Given the success of applying BERT to different ranking tasks, is it possible to improve the quality of target apps selection using a BERT-based ranker?
\item \textbf{RQ2}: Is BERT able to leverage its knowledge of natural language to select a more diverse set of apps, compared to other IR models? Which apps benefit most from this?
\end{itemize}

We provide experimental results on the UniMobile dataset for a BERT-based ranker to answer these research questions. The contributions of this work are as follows:

\begin{itemize}[leftmargin=*]
\item We show that using a BERT-based ranker for the target app selection task improves performance compared to the previous state-of-the-art baselines.
\item Our analysis reveals more diverse results of the BERT-based ranker compared with neural and IR baselines. We show that resource-scarce apps such as Contacts and File Manager benefit the most from this ranker's diversity.
\end{itemize}

Our results suggest that pre-trained transformers can be applied to resource selection tasks, especially when a small amount of data is available about specific resources. Such models are able to model resources based on the queries submitted to them (rather than their content) more effectively. Our experiments reveal that the BERT-based ranker is less biased towards selecting the most popular app (i.e., Google Search) and instead outperforms other models in predicting less popular apps by a large margin.

\section{Pre-Trained Transformers for Resource Selection}
\label{sec:model}
Prior work demonstrates that using BERT-based rankers can achieve state-of-the-art results on various information retrieval tasks such as passage ranking ~\citep{passageReRanking}, document ranking ~\citep{MultiStage}, and question answering ~\citep{Yang2019EndtoEndOQ}. We follow a common BERT-based ranking approach introduced by ~\citet{passageReRanking} in which the input vector is a query--document pair. This approach follows the BERT's standard input format and uses a concatenated string of a [CLS] token,  query, [SEP] token, document, and another [SEP] token.

As mentioned in Section~\ref{sec:introduction}, a mobile device environment is counted as uncooperative in the sense that apps do not share their representations. To overcome this problem and make resource selection more similar to the document ranking task, we use a random combination of all queries related to an app in the training set to build a document representation for each app. In addition, different queries are separated using a comma to make the representation more similar to a natural language text. Since some apps have many queries and BERT has a limitation on the number of tokens, if the input vector length is longer than the model's maximum length, we truncate the document, but the query remains unchanged.

We use the BERT-base, uncased, pre-trained model with an added single linear classification layer on top of the $[CLS]$ output vector, utilizing the model for resource selection. In this case, Each input's $[CLS]$ output vector would be fed to the classification layer, and both the pre-trained BERT model and the additional untrained classification layer are tuned on training samples.

We used the train-test splits used by ~\citet{UniMobile}. The fine-tuning process is performed using the top-$10$ selection of BM25 results for each query in the training dataset. We used a learning rate of 2e-5 with no warmup, batch size 16, 2 epochs, and BERT's maximum token limitation of 256 in our experiments. After fine-tuning the BERT model, we perform the testing process on queries from the held-out test set.

\begin{table*}
    \centering
    \vspace{-2mm}
    \caption{Performance comparison with baselines on UniMobile-Q and UniMobile-T. The superscript * denotes significant differences compared to all the baselines.}
    \vspace{-2mm}
    \label{tab:results}
    \resizebox{17.9cm}{!}{%
    \begin{tabular}{lccccccccccc}
    \toprule
     \multirow{2}{*}{\textbf{Method}} & \multicolumn{5}{c}{\textbf{UniMobile-Q Dataset}} && \multicolumn{5}{c}{\textbf{UniMobile-T Dataset}} \\
     \cmidrule{2-6} \cmidrule{8-12} 
     & MRR & P@1 & nDCG@1 & nDCG@3 & nDCG@5 && MRR & P@1 & nDCG@1 & nDCG@3 & nDCG@5\\
    \midrule
    \textbf{StaticRanker} & 0.6485 & 0.5293 & 0.4031 & 0.4501 & 0.5144 && 0.6718 & 0.5507 & 0.4247 & 0.4853 & 0.5446 \\
    \textbf{QueryLM} & 0.5867 & 0.3803 & 0.3068 & 0.4676 & 0.5508 && 0.5178 & 0.3272 & 0.2619 & 0.3716 & 0.4503 \\ %
    \textbf{BM25} & 0.7523 & 0.6233 & 0.4915 & 0.6298 & 0.6859 && 0.6780 & 0.5244 & 0.4101 & 0.5392 & 0.5992 \\ %
    \textbf{BM25-QE} & 0.6948 & 0.5177 & 0.4116 & 0.5909 & 0.6498 && 0.6256 & 0.4276 & 0.3312 & 0.5015 & 0.5704 \\ %
    \textbf{k-NN} & 0.7373 & 0.6031 & 0.4794 & 0.6091 & 0.6633 && 0.6879 & 0.5414 & 0.4287 & 0.5413 & 0.6003 \\
    \textbf{k-NN-AWE} & 0.7420 & 0.6081 & 0.4842 & 0.6156 & 0.6682 && 0.6984 & 0.5551 & 0.4407 & 0.5560 & 0.6117 \\
    \textbf{LambdaMART} & 0.7313 & 0.6127 & 0.4864 & 0.6110 & 0.6426 && 0.6749 & 0.5469 & 0.4323 & 0.5419 & 0.5704 \\
    \textbf{NTAS} & 0.7661 & 0.6285 & 0.5012 & 0.6364 & 0.7018 & & 0.7192 & 0.5661 & 0.4709 & 0.5941 & 0.6471 \\
    \midrule
    \textbf{BERT} & \textbf{0.7685} & \textbf{0.6383*} & \textbf{0.5246*} & \textbf{0.6520*} & \textbf{0.7021} && \textbf{0.7395*} & \textbf{0.5960*} & \textbf{0.4827*} & \textbf{0.6154*} & \textbf{0.6705*} \\
    \bottomrule
    \end{tabular}
    }
    \vspace{-3mm}
\end{table*}

\section{Experiment}

\subsection{Experimental Setup}

\partitle{Dataset.}
We evaluate our approach using the UniMobile dataset, which was collected and published by ~\citet{UniMobile}. They used crowdsourcing to collect query and app pairs based on pre-defined tasks. There are $206$ unique tasks with the aim of covering various search categories. The workers would first read the task description and then assume that they want to perform the task using their own mobile device. They also were asked to select one or more most helpful apps according to their query after typing their query. 

We follow the related work in splitting the data~\cite{UniMobile,Situ,DBLP:journals/corr/abs-2101-03394}: (1) \textit{Uni\-Mobile-Q} in which queries are selected randomly for the train, validation, and test (2) \textit{UniMobile-T} in which tasks are selected randomly. Therefore, while in UniMobile-Q the queries of the same search tasks is seen in the training set, in UniMobile-T the tasks are completely unseen in the test set. We also follow the related work ~\citep{UniMobile} and measure the performance in terms of the following metrics: MRR, P@1, nDCG@1, nDCG@3, and nDCG@5. Furthermore, we follow the related work in assigning a score of 2 to the \textit{first} relevant app and 1 to the rest of relevant apps, to differentiate between a model that is able to rank the first relevant app higher and a model that is not. For MRR and P@1, apps with score 2 or 1 are considered relevant.

\partitle{Baselines.}
We compare the performance of the BERT-based ranker with a number of IR and machine learning methods, as listed below:

\begin{itemize}[leftmargin=*]
\item StaticRanker: A query-independent model which uses a ranked list of most popular apps on the training set.

\item QueryLM, BM25, BM25-QE: As mentioned in Section~\ref{sec:model}, we build used all relevant queries from the training set as a single document representation for each app. These methods use the mentioned document representations as a corpus to find the best app for each query. BM25 parameters were tuned on the validation set and are as follows: $k1 = 1.5$ and $b = 0$.

\item k-NN, k-NN-AWE: The cosine similarity between TF-IDF and the average word embedding (AWE) vectors of the queries were used in k-NN and k-NN-AWE, respectively. Then the apps related to the nearest queries were used to produce the app ranking.

\item LambdaMART: For every query-app pair, the scores obtained by BM25, k-NN, and k-NN-AWE are used as features to train LambdaMART.  All irrelevant apps are considered as negative samples for every query.

\item NTAS: A neural model approach, designed by ~\citep{UniMobile} for the target apps selection task in which the query and apps representations and their related score are learned using a network. We use the NTAS1-pairwise model due to its superior performance compared to the other variants.

\end{itemize}

\partitle{Significance testing.}
We determined the statistically significant differences using the two-tailed paired t-test with Bonferroni correction at a $95\%$ confidence interval ($p < 0.05$). The performance of the BERT-based ranker is marked as significant if its test indicates significance in comparison with all baseline approaches.

\vspace{-2mm}
\subsection{Results and Discussion}

\partitle{Performance comparison.}
The last row of Table~\ref{tab:results} shows our BERT-based ranker results, indicating improvement over previous baselines.
These experiments, answer our \textbf{RQ1} and show the BERT-based ranker outperforms all the baselines in terms of all evaluation metrics. Moreover, there is a higher performance improvement on the UniMobile-T data split, leading to significant improvement compared to all baselines in terms of all evaluation metrics.
There is no query belonging to the same task in training and test sets in this split, so more remarkable improvement suggests that the BERT-based ranker can capture related query similarities and then use generalization to consider similar queries for unseen tasks. This is evident when we compare BERT's performance with term-matching methods such as BM25, where we observe a 12\% relative improvement in terms of nDCG@5. It is worth mentioning that the BERT-based ranker also exhibits a larger margin of improvement compared to models based on word embeddings, where it achieves a relative improvement of 10\% and 4\% in terms of nDCG@5 compared to k-NN-AWE and NTAS, respectively.

On the other hand, on UniMobile-Q we see that the BERT-based ranker achieves significant improvements only in terms of P@1, nDCG@1, and nDCG@3, indicating the BERT-based ranker's ability to enhance the top of the ranking more effectively. This suggests that in cases where there is a high rate of term-match between the user's query and the resources, the BERT-based ranker still outperforms the baselines, but to a lower extent, suggesting that BERT's strength lies in semantic and structural properties of queries.

\begin{figure}
    \centering
    \vspace{-7mm}
    \subfloat[UniMobile-Q]{\includegraphics[trim=0.0cm 2.22cm 0.0cm 0.0cm, clip,width=\columnwidth]{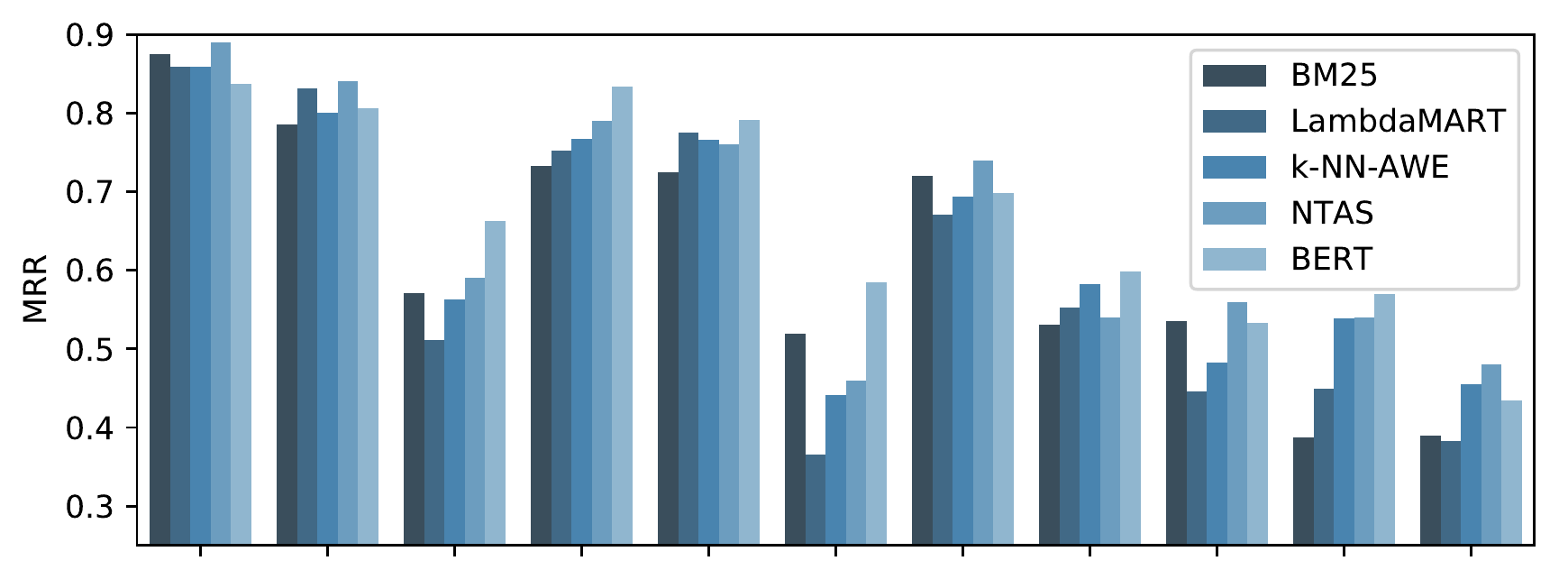}\label{fig:app_q}}
    
    \vspace{-3mm}
    \subfloat[UniMobile-T]{\includegraphics[width=\columnwidth]{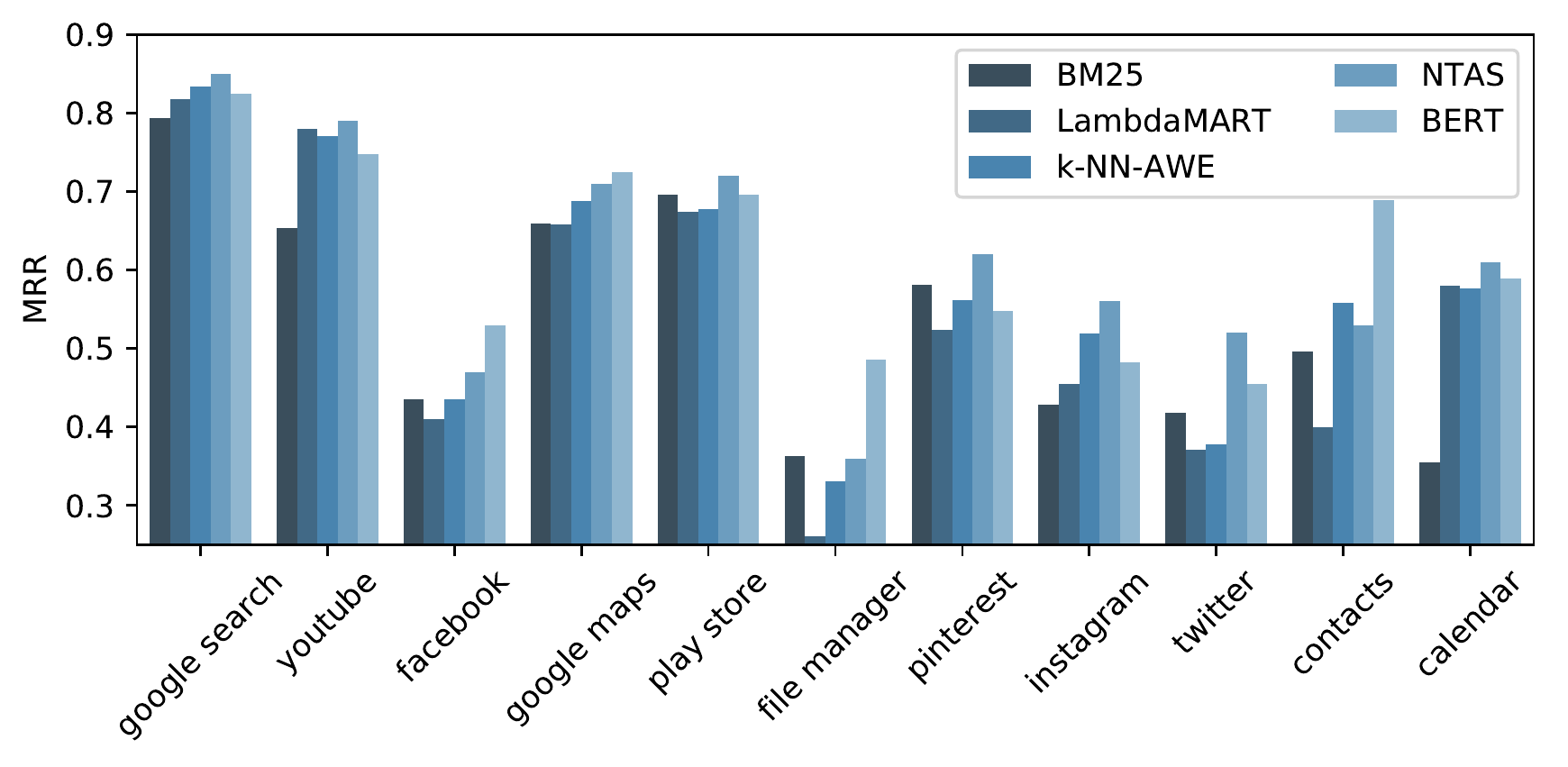}\label{fig:app_t}}
    \vspace{-3mm}
    \caption{Performance comparison based on different apps.}
    \vspace{-4mm}
    \label{fig:app_based}
\end{figure}

\partitle{Diversity.}
We compute the mean performance of the queries targeted to a specific app and plot each app's result in Figure~\ref{fig:app_based} to investigate our \textbf{RQ2}. We only include the comparison of the best NTAS model and other baseline results with the BERT-based ranker in terms of MRR\@. We see that all models show lower performance on UniMobile-T compared to UniMobile-Q. This pattern is expected as the queries for the same task are observed in the training data; therefore, the model can learn better in UniMobile-Q (except for Contacts and Calendar).

Even though the margin is small, we see that the BERT-based ranker performs worse for apps with more generic queries such as Google Search and YouTube. On the other hand, we see that the BERT-based ranker can outperform other models for apps with more personal queries that need more semantic understanding. For example, we see that the BERT-based ranker outperforms other models for Facebook queries by a large margin.

Among the selected apps, we see a larger margin when comparing the queries submitted to Contacts, where understanding the semantics and structure of queries can be more important. For example, the model can use the fact that the query is a person's first name to select the Contacts app as query's target. According to ~\cite{UniMobile}, Contacts and Facebook have lower query overlap~\cite{DBLP:journals/tweb/ChurchSCB07} when compared to apps such as Google Search and YouTube. This could explain the different behavior we observe from the BERT-based ranker compared to other models, suggesting that the BERT-based ranker is able to leverage its knowledge of language and structure, together with the high-level understanding of query intent in the process of resource selection.

Moreover, we observe that BERT's superiority in incorporating query semantics and structure is magnified on UniMobile-T where queries of the same task are not included in the training data. This indicates that other models depend more on the exact matching of terms, while BERT can leverage its knowledge of language. Specifically, compared to NTAS, we see that the BERT-based ranker exhibits a large improvement for the Contacts and File Manager apps on the task-based data split. 

To give an overview of the app-based performance of BERT for all apps, we extend the analysis of Figure~\ref{fig:app_based} and compute an average app-based performance in terms of MRR for all apps (i.e., macro-averaging MRR for all the apps). We see that BERT's macro-average MRR on UniMobile-Q is $0.4356$, which is higher than BM25's, with $0.4229$. The measure margin is also higher on UniMobile-T, which is $0.3991$ and $0.3666$ for BERT and BM25, respectively. This measure can show the effect of apps with lower queries. The higher macro-average MRR (as opposed to micro-average MRR in Table~\ref{tab:results}) indicates effectiveness of BERT in predicting not only high-resource apps, but also apps that have less queries.

\partitle{Error analysis.}
In this experiment, we aim to compare the BERT-based ranker's performance with a term-matching method like BM25. Our goal is to analyze the cases where exact-term matching cannot help identify the resource and see how BERT's semantic and language knowledge can benefit effective resource selection. As we saw in Figure~\ref{fig:app_based}, BERT exhibited an improved performance compared to all baselines for apps that had fewer training queries or less query similarity. This already suggests that BERT leverages the semantic similarity of queries and, perhaps in many cases, the structure of queries. Another challenge here is the dominance of apps such as Google Search, which are the target apps for nearly half of the submitted queries. Apart from having imbalanced labels, apps like Google Search impose another challenge: being the app-for-all -- being the target app for a diverse set of tasks. Therefore, models can easily confuse Google Search with almost any other app. 

Therefore, in Table~\ref{tab:analysis} we list the rate of cases where each model confuses the true label with Google Search. To compute each row, we count the number of times that the best app in the gold data is the row's app (e.g., YouTube), but is instead labeled as Google Search. We see in the table that BERT exhibits a higher average error rate for the UniMobile-T (0.4540) compared to UniMobile-Q (0.3568), indicating that overlapping queries for the same task increase the model's capability of modeling the resources more effectively. Moreover, we see that, on average, BERT has a much lower error rate compared to BM25 on both datasets. The fact that BM25 has a higher error rate on UniMobile-Q suggests more query overlap in the training set increases the risk of biasing towards Google Search. However, we see that BERT is able to circumvent this problem, as it shows a lower average error rate. This could be due to its ability to understand the natural language more effectively and its generalization.

Taking a closer look at the error rates, we see that the highest relative difference between the two models occurs for the Contacts app on both datasets. In particular, we see that BERT's error rate on UniMobile-Q is nearly a quarter of BM25. Also, we see that the error rate on UniMobile-T is less than half on UniMobile-T. We also see a high difference for Facebook. Considering the low query overlap~\cite{UniMobile} and the fact that the queries submitted to these apps are very personal (e.g., friend and family names), we see that BERT is successfully distinguishing these apps from Google Search, perhaps by taking into account their structure and nature (i.e., being short and proper nouns). Notice that, even though we observe significant differences in terms of Google Search confusion error rate on UniMobile-Q in Table~\ref{tab:analysis}, we see more performance gain on UniMobile-T in Table~\ref{tab:results} and Figure~\ref{fig:app_based}. It is worth mentioning that these results do not necessarily correlate since a model can have less confusion with Google Search, yet at the same time confuse an app with another app.

\begin{table}
    \centering
    \vspace{-2mm}
    \caption{Error analysis on different apps. The numbers show the rate of cases where the Google Search app is chosen instead of a more specific app named in the column. Lower numbers show less error, hence more desired.}
    \vspace{-2mm}
    \label{tab:analysis}
    \resizebox{8.8cm}{!}{%
    \begin{tabular}{lccccccc}
    \toprule
     \multirow{2}{*}{\textbf{App}} & \multicolumn{3}{c}{\textbf{UniMobile-Q Dataset}} && \multicolumn{3}{c}{\textbf{UniMobile-T Dataset}} \\
     \cmidrule{2-4} \cmidrule{6-8} 
     & BERT & BM25 & $\Delta$(\%) && BERT & BM25 & $\Delta$(\%) \\
    \midrule
    \textbf{google maps}    & 0.3511	& 0.7400    & 111	    && 0.6440	& 0.7414    & 15 \\
\textbf{instagram}          & 0.4848	& 0.5530    & 14	        && 0.5239	& 0.5181    & 1 \\
\textbf{facebook}           & 0.3836	& 0.5807    & 51        && 0.4993	& 0.5867    & 17 \\
\textbf{pinterest}          & 0.5750	& 0.7795    & 36        && 0.5977	& 0.4315    & 28 \\
\textbf{calendar}           & 0.2461	& 0.3403    & 38        && 0.2690	& 0.3820    & 42 \\
\textbf{play store}         & 0.3279	& 0.6262    & 91        && 0.4651	& 0.6429    & 38 \\
\textbf{file manager}       & 0.2628	& 0.4038    & 54	        && 0.3772	& 0.5482    & 45 \\
\textbf{youtube}            & 0.4436	& 0.5998    & 35        && 0.4891	& 0.6286    & 28 \\
\textbf{contacts}           & 0.0811	& 0.4023    & 396        && 0.2100	& 0.4731    & 125 \\
\textbf{twitter}            & 0.4121	& 0.4118    & 0        && 0.4647	& 0.3238    & 30 \\
\midrule    
\textbf{average}            & 0.3568    & 0.5437    & 52        && 0.4540   & 0.5276    & 16 \\
    \bottomrule
    \end{tabular}
    }
    \vspace{-5mm}
\end{table}

\section{Conclusions and Future Work}
This paper investigates the benefits of using a vanilla BERT-based ranker for resource selection in a unified mobile search environment. In particular, we chose the target app selection task, as it imposes novel challenges such as having heterogeneous data in an uncooperative environment. Also, the dominance of certain resources such as Google Search imposes further challenge. We study how BERT performs with respect to these challenges, compared to other neural and IR models.

Our experiments on the UniMobile dataset show that the BERT-based ranker outperforms state-of-the-art models. Our observations suggest that the BERT-based ranker learns semantics and the structure of queries better, leading the final ranking to be more diverse. We showed that using a BERT-based ranker is especially effective for the apps with more personal queries such as Facebook and Contacts. Moreover, we observed that the BERT-based ranker is less biased towards the most popular resource. 

Our preliminary analyses suggest that BERT leverages the semantics and structure queries in prediction. We plan to investigate this further by a detailed analysis of different layers of the model in the process of fine-tuning~\cite{DBLP:conf/cikm/AkenWLG19}. As mentioned, in this work, we sampled the queries associated to each app, and in case it did not fit within the BERT's token-size limitation, we truncated it. We plan to extend our method to accommodate a better representation for each app, inspired by related work on context-aware term weighting~\cite{DBLP:conf/www/DaiC20}. Finally, we plan to examine the effect of using other app metadata to enhance resource descriptions of resource-scarce apps.

\newpage
\bibliographystyle{ACM-Reference-Format}
\bibliography{refs}


\begin{thebibliography}{17}


\ifx \showCODEN    \undefined \def \showCODEN     #1{\unskip}     \fi
\ifx \showDOI      \undefined \def \showDOI       #1{#1}\fi
\ifx \showISBNx    \undefined \def \showISBNx     #1{\unskip}     \fi
\ifx \showISBNxiii \undefined \def \showISBNxiii  #1{\unskip}     \fi
\ifx \showISSN     \undefined \def \showISSN      #1{\unskip}     \fi
\ifx \showLCCN     \undefined \def \showLCCN      #1{\unskip}     \fi
\ifx \shownote     \undefined \def \shownote      #1{#1}          \fi
\ifx \showarticletitle \undefined \def \showarticletitle #1{#1}   \fi
\ifx \showURL      \undefined \def \showURL       {\relax}        \fi
\providecommand\bibfield[2]{#2}
\providecommand\bibinfo[2]{#2}
\providecommand\natexlab[1]{#1}
\providecommand\showeprint[2][]{arXiv:#2}

\bibitem[\protect\citeauthoryear{Aliannejadi, Zamani, Crestani, and
  Croft}{Aliannejadi et~al\mbox{.}}{2018a}]%
        {Situ}
\bibfield{author}{\bibinfo{person}{Mohammad Aliannejadi},
  \bibinfo{person}{Hamed Zamani}, \bibinfo{person}{Fabio Crestani}, {and}
  \bibinfo{person}{W.~Bruce Croft}.} \bibinfo{year}{2018}\natexlab{a}.
\newblock \showarticletitle{In Situ and Context-Aware Target Apps Selection for
  Unified Mobile Search}. In \bibinfo{booktitle}{\emph{CIKM}}.
  \bibinfo{pages}{1383–1392}.
\newblock


\bibitem[\protect\citeauthoryear{Aliannejadi, Zamani, Crestani, and
  Croft}{Aliannejadi et~al\mbox{.}}{2018b}]%
        {UniMobile}
\bibfield{author}{\bibinfo{person}{Mohammad Aliannejadi},
  \bibinfo{person}{Hamed Zamani}, \bibinfo{person}{Fabio Crestani}, {and}
  \bibinfo{person}{W.~Bruce Croft}.} \bibinfo{year}{2018}\natexlab{b}.
\newblock \showarticletitle{Target Apps Selection: Towards a Unified Search
  Framework for Mobile Devices}. In \bibinfo{booktitle}{\emph{SIGIR}}.
  \bibinfo{pages}{215–224}.
\newblock


\bibitem[\protect\citeauthoryear{Aliannejadi, Zamani, Crestani, and
  Croft}{Aliannejadi et~al\mbox{.}}{2021}]%
        {DBLP:journals/corr/abs-2101-03394}
\bibfield{author}{\bibinfo{person}{Mohammad Aliannejadi},
  \bibinfo{person}{Hamed Zamani}, \bibinfo{person}{Fabio Crestani}, {and}
  \bibinfo{person}{W.~Bruce Croft}.} \bibinfo{year}{2021}\natexlab{}.
\newblock \showarticletitle{Context-Aware Target Apps Selection and
  Recommendation for Enhancing Personal Mobile Assistants}.
\newblock \bibinfo{journal}{\emph{CoRR}}  \bibinfo{volume}{abs/2101.03394}
  (\bibinfo{year}{2021}).
\newblock


\bibitem[\protect\citeauthoryear{Arguello, Diaz, and Callan}{Arguello
  et~al\mbox{.}}{2011}]%
        {DBLP:conf/cikm/ArguelloDC11}
\bibfield{author}{\bibinfo{person}{Jaime Arguello}, \bibinfo{person}{Fernando
  Diaz}, {and} \bibinfo{person}{Jamie Callan}.}
  \bibinfo{year}{2011}\natexlab{}.
\newblock \showarticletitle{Learning to aggregate vertical results into web
  search results}. In \bibinfo{booktitle}{\emph{{CIKM}}}.
  \bibinfo{publisher}{{ACM}}, \bibinfo{pages}{201--210}.
\newblock


\bibitem[\protect\citeauthoryear{Arguello, Diaz, and Paiement}{Arguello
  et~al\mbox{.}}{2010}]%
        {DBLP:conf/sigir/ArguelloDP10}
\bibfield{author}{\bibinfo{person}{Jaime Arguello}, \bibinfo{person}{Fernando
  Diaz}, {and} \bibinfo{person}{Jean{-}Fran{\c{c}}ois Paiement}.}
  \bibinfo{year}{2010}\natexlab{}.
\newblock \showarticletitle{Vertical selection in the presence of unlabeled
  verticals}. In \bibinfo{booktitle}{\emph{{SIGIR}}}.
  \bibinfo{publisher}{{ACM}}, \bibinfo{pages}{691--698}.
\newblock


\bibitem[\protect\citeauthoryear{Callan and Connell}{Callan and
  Connell}{2001}]%
        {CallanConnell}
\bibfield{author}{\bibinfo{person}{Jamie Callan} {and}
  \bibinfo{person}{Margaret Connell}.} \bibinfo{year}{2001}\natexlab{}.
\newblock \showarticletitle{Query-Based Sampling of Text Databases}.
\newblock \bibinfo{journal}{\emph{ACM Trans. Inf. Syst.}} \bibinfo{volume}{19},
  \bibinfo{number}{2} (\bibinfo{year}{2001}), \bibinfo{pages}{97–130}.
\newblock


\bibitem[\protect\citeauthoryear{Church, Smyth, Cotter, and Bradley}{Church
  et~al\mbox{.}}{2007}]%
        {DBLP:journals/tweb/ChurchSCB07}
\bibfield{author}{\bibinfo{person}{Karen Church}, \bibinfo{person}{Barry
  Smyth}, \bibinfo{person}{Paul Cotter}, {and} \bibinfo{person}{Keith
  Bradley}.} \bibinfo{year}{2007}\natexlab{}.
\newblock \showarticletitle{Mobile information access: {A} study of emerging
  search behavior on the mobile Internet}.
\newblock \bibinfo{journal}{\emph{{ACM} Trans. Web}} \bibinfo{volume}{1},
  \bibinfo{number}{1} (\bibinfo{year}{2007}), \bibinfo{pages}{4}.
\newblock
\urldef\tempurl%
\url{https://doi.org/10.1145/1232722.1232726}
\showDOI{\tempurl}


\bibitem[\protect\citeauthoryear{Crestani, Mizzaro, and Scagnetto}{Crestani
  et~al\mbox{.}}{2017}]%
        {DBLP:series/sbcs/CrestaniMS17}
\bibfield{author}{\bibinfo{person}{Fabio Crestani}, \bibinfo{person}{Stefano
  Mizzaro}, {and} \bibinfo{person}{Ivan Scagnetto}.}
  \bibinfo{year}{2017}\natexlab{}.
\newblock \bibinfo{booktitle}{\emph{Mobile Information Retrieval}}.
\newblock \bibinfo{publisher}{Springer}.
\newblock


\bibitem[\protect\citeauthoryear{Dai and Callan}{Dai and Callan}{2020}]%
        {DBLP:conf/www/DaiC20}
\bibfield{author}{\bibinfo{person}{Zhuyun Dai} {and} \bibinfo{person}{Jamie
  Callan}.} \bibinfo{year}{2020}\natexlab{}.
\newblock \showarticletitle{Context-Aware Document Term Weighting for Ad-Hoc
  Search}. In \bibinfo{booktitle}{\emph{{WWW}}}. \bibinfo{publisher}{{ACM} /
  {IW3C2}}, \bibinfo{pages}{1897--1907}.
\newblock


\bibitem[\protect\citeauthoryear{Devlin, Chang, Lee, and Toutanova}{Devlin
  et~al\mbox{.}}{2019}]%
        {BERT}
\bibfield{author}{\bibinfo{person}{Jacob Devlin}, \bibinfo{person}{Ming-Wei
  Chang}, \bibinfo{person}{Kenton Lee}, {and} \bibinfo{person}{Kristina
  Toutanova}.} \bibinfo{year}{2019}\natexlab{}.
\newblock \showarticletitle{{BERT}: Pre-training of Deep Bidirectional
  Transformers for Language Understanding}. In
  \bibinfo{booktitle}{\emph{NAACL}}. \bibinfo{pages}{4171--4186}.
\newblock


\bibitem[\protect\citeauthoryear{MacAvaney, Yates, Cohan, and
  Goharian}{MacAvaney et~al\mbox{.}}{2019}]%
        {CEDR}
\bibfield{author}{\bibinfo{person}{Sean MacAvaney}, \bibinfo{person}{Andrew
  Yates}, \bibinfo{person}{Arman Cohan}, {and} \bibinfo{person}{Nazli
  Goharian}.} \bibinfo{year}{2019}\natexlab{}.
\newblock \showarticletitle{CEDR: Contextualized Embeddings for Document
  Ranking}. In \bibinfo{booktitle}{\emph{SIGIR}}. \bibinfo{pages}{1101–1104}.
\newblock


\bibitem[\protect\citeauthoryear{Nogueira and Cho}{Nogueira and Cho}{2019}]%
        {passageReRanking}
\bibfield{author}{\bibinfo{person}{Rodrigo Nogueira} {and}
  \bibinfo{person}{Kyunghyun Cho}.} \bibinfo{year}{2019}\natexlab{}.
\newblock \showarticletitle{Passage Re-ranking with BERT}.
\newblock \bibinfo{journal}{\emph{arXiv preprint arXiv:1901.04085}}
  (\bibinfo{year}{2019}).
\newblock


\bibitem[\protect\citeauthoryear{Nogueira, Yang, Cho, and Lin}{Nogueira
  et~al\mbox{.}}{2019}]%
        {MultiStage}
\bibfield{author}{\bibinfo{person}{Rodrigo Nogueira}, \bibinfo{person}{Wei
  Yang}, \bibinfo{person}{Kyunghyun Cho}, {and} \bibinfo{person}{Jimmy Lin}.}
  \bibinfo{year}{2019}\natexlab{}.
\newblock \showarticletitle{Multi-stage document ranking with BERT}.
\newblock \bibinfo{journal}{\emph{arXiv preprint arXiv:1910.14424}}
  (\bibinfo{year}{2019}).
\newblock


\bibitem[\protect\citeauthoryear{Park, Liu, Zhai, and Wang}{Park
  et~al\mbox{.}}{2015}]%
        {Park}
\bibfield{author}{\bibinfo{person}{Dae~Hoon Park}, \bibinfo{person}{Mengwen
  Liu}, \bibinfo{person}{ChengXiang Zhai}, {and} \bibinfo{person}{Haohong
  Wang}.} \bibinfo{year}{2015}\natexlab{}.
\newblock \showarticletitle{Leveraging User Reviews to Improve Accuracy for
  Mobile App Retrieval}. In \bibinfo{booktitle}{\emph{SIGIR}}.
  \bibinfo{pages}{533–542}.
\newblock


\bibitem[\protect\citeauthoryear{Shokouhi and Si}{Shokouhi and Si}{2011}]%
        {Shokouhi}
\bibfield{author}{\bibinfo{person}{Milad Shokouhi} {and} \bibinfo{person}{Luo
  Si}.} \bibinfo{year}{2011}\natexlab{}.
\newblock \showarticletitle{Federated Search}.
\newblock \bibinfo{journal}{\emph{Found. Trends Inf. Retr.}}
  \bibinfo{volume}{5}, \bibinfo{number}{1} (\bibinfo{year}{2011}),
  \bibinfo{pages}{1–102}.
\newblock


\bibitem[\protect\citeauthoryear{van Aken, Winter, L{\"{o}}ser, and Gers}{van
  Aken et~al\mbox{.}}{2019}]%
        {DBLP:conf/cikm/AkenWLG19}
\bibfield{author}{\bibinfo{person}{Betty van Aken}, \bibinfo{person}{Benjamin
  Winter}, \bibinfo{person}{Alexander L{\"{o}}ser}, {and}
  \bibinfo{person}{Felix~A. Gers}.} \bibinfo{year}{2019}\natexlab{}.
\newblock \showarticletitle{How Does {BERT} Answer Questions?: {A} Layer-Wise
  Analysis of Transformer Representations}. In
  \bibinfo{booktitle}{\emph{{CIKM}}}. \bibinfo{publisher}{{ACM}},
  \bibinfo{pages}{1823--1832}.
\newblock


\bibitem[\protect\citeauthoryear{Yang, Xie, Lin, Li, Tan, Xiong, Li, and
  Lin}{Yang et~al\mbox{.}}{2019}]%
        {Yang2019EndtoEndOQ}
\bibfield{author}{\bibinfo{person}{Wei Yang}, \bibinfo{person}{Yuqing Xie},
  \bibinfo{person}{Aileen Lin}, \bibinfo{person}{Xingyu Li},
  \bibinfo{person}{Luchen Tan}, \bibinfo{person}{Kun Xiong},
  \bibinfo{person}{Ming Li}, {and} \bibinfo{person}{Jimmy Lin}.}
  \bibinfo{year}{2019}\natexlab{}.
\newblock \showarticletitle{End-to-End Open-Domain Question Answering with
  BERTserini}. In \bibinfo{booktitle}{\emph{NAACL-HLT}}.
\newblock


\end{thebibliography}

\end{document}